\documentclass[openany]{nature}
\usepackage[T1]{fontenc}
\usepackage[left]{lineno}
\usepackage{amsthm,amsmath,amssymb}
\usepackage{mathrsfs}
\usepackage{overpic}
\usepackage{multirow}
\usepackage[T1]{fontenc}
\usepackage{graphicx}
\usepackage{pdflscape}
\usepackage{hyperref}
\usepackage{threeparttable}
\usepackage{threeparttablex}
\usepackage{xcolor}
\usepackage{ulem}
\usepackage{cancel}
\usepackage[figuresleft]{rotating}
\usepackage{caption}
\usepackage{tablefootnote}
\usepackage{hyperref}
\usepackage{graphicx}
\usepackage{longtable}
\usepackage{supertabular}
\usepackage{float} 
\usepackage{authblk}
\usepackage{multibib}
\usepackage{subfiles}
\usepackage{enumerate}
\usepackage{bm}
\usepackage{ulem}
\usepackage{comment}
\usepackage{subfloat}
\usepackage{subcaption}
\usepackage{url}
\usepackage{multirow}
\usepackage{makecell}
\usepackage{geometry}
\makeatletter
 
\def\emph#1 {\textit{ #1 } }

\let\saved@includegraphics\includegraphics
\AtBeginDocument{\let\includegraphics\saved@includegraphics}
\renewenvironment*{figure}{\@float{figure}}{\end@float}
\makeatother

\providecommand{\mbh}{M_{\bullet}}
\providecommand{\jbh}{J_{\bullet}}
\providecommand{\Pbz}{P_{\rm BZ}}

\providecommand{\Rh}{R_{\bullet}}
\title{Gravitational Wave Evidence of Spin Energy Extraction from Black Holes}
\author{
Shu-Xu Yi$^{1,2}$\thanks{Email: sxyi@ihep.ac.cn}, Tian-Yong Cao $^{1,2}$, Shuang-Nan Zhang$^{1,2}$\thanks{Email: zhangsn@ihep.ac.cn}, Hua Feng$^{1,2}$}

\date{March 2025}    
\begin{document}
\maketitle
\begin{affiliations}
\item State Key Laboratory of Particle Astrophysics, Institute of High Energy Physics, Chinese Academy of Sciences, Beijing 100049, China.
\item University of Chinese Academy of Sciences, Chinese Academy of Sciences, Beijing 100049, China.
\end{affiliations}
\begin{abstract}
Relativistic jet is a key phenomenon in energetic astrophysical objects, yet its energy resource remains a mystery. Some attribute it to accretion energy, while others, more interestingly, to magnetic extraction of BH rotational energy. There is no decisive observational distinction yet. We argue that if the latter scenario holds, the spin-up of the BH via the natal accretion of its progenitor stellar matter, and spin-down via magnetic extraction of its rotational energy can reach a dynamic balance. In the case of magnetically arrested disk (MAD) accretion, the BH spin converges to an equilibrium value that depends solely on the physics of the accretion flow near the BH event horizon. On the other hand, if jet is powered by accretion, no such universal equilibrium spin should be expected. Therefore, a population of stellar-mass BHs with a universal spin is a strong signature of BH rotational energy extraction. Testing against the 4th Gravitational Wave (GW) Transients Catalogue (GWTC-4.0), we find a statistically robust dominant population where second-born BH spins are narrowly centred at $\sim0.05$. These findings provide strong new evidence for BH spin energy extraction, which is scarcely explicable in the accretion-powered scenario.

\end{abstract}
\section{Main text}
The progenitors of type-II gamma-ray bursts (GRBs) are believed to be collapsing massive stars \cite{2006NatPh...2..116G,2008AIPC.1000...11B,2009A&A...496..585G,2013ApJ...764..179B,2012ApJ...759..107K}. The core of the massive star first collapses into a prompt BH, then a large fraction of the remaining stellar material will be accreted onto this BH overall a timescale of 10-1000 seconds, with a hyperaccretion rate \cite{2019ApJ...871..117S,2019ApJ...878..142W,2023ApJ...956..100F}. In this process, an accretion disk and the magnetic field sustained by it are crucial for launching of a relativistic jet of the GRB \cite{2006MNRAS.368.1561M,2022A&A...668A..66J}. In the theory of Blandford \& Znajek\cite{1977MNRAS.179..433B}, the magnetic field lines threading the event horizon of the BH can extract the BH's rotational energy and drive the relativistic jet. Indeed, many GRBs have been observed with evidence of jets possessing significant magnetic field ($\sim$10 G) \cite{2023arXiv231007205Y,2024ApJ...973..113T,2024ApJ...972....1L,2024MNRAS.529L..67D}, which supports the Blandford-Znajek (BZ) scenario at the central engine. Therefore, the accretion disk has a two-fold impact on the spin evolution of the BH. On one hand, matter falling from the inner edge of the accretion disk adds to the rotational energy (and the angular momentum) of the BH; on the other hand, the magnetic field sustained by the disk extracts the BH's spin energy and angular momentum. If the accretion disk is in the Magnetically Arrested Disk (MAD) state, the magnetic pressure is comparable to the ram pressure of the inflowing matter at the event horizon \cite{1974Ap&SS..28...45B,1976Ap&SS..42..401B,2003ApJ...592.1042I,2003PASJ...55L..69N}. Consequently, the magnetic field strength at the event horizon is positively correlated with the accretion rate. Therefore, in the process of hyper accretion, these two competing process may reach a balance, such that the BH spin attains an equilibrium value. Recent numerical simulations have shown that the equilibrium spin of this process is small ($\chi<0.1$), if the MAD state is accomplished at the onset of accretion \cite{2024ApJ...961..212J}. As we shall demonstrate, the equilibrium spin can be derived analytically; its value is independent of the BH's initial spin, mass and accretion rate history, but depends solely on the physics of the accretion flow in the vicinity of the BH event horizon.  

\begin{figure}
    \centering
    \includegraphics[width=0.5\textwidth]{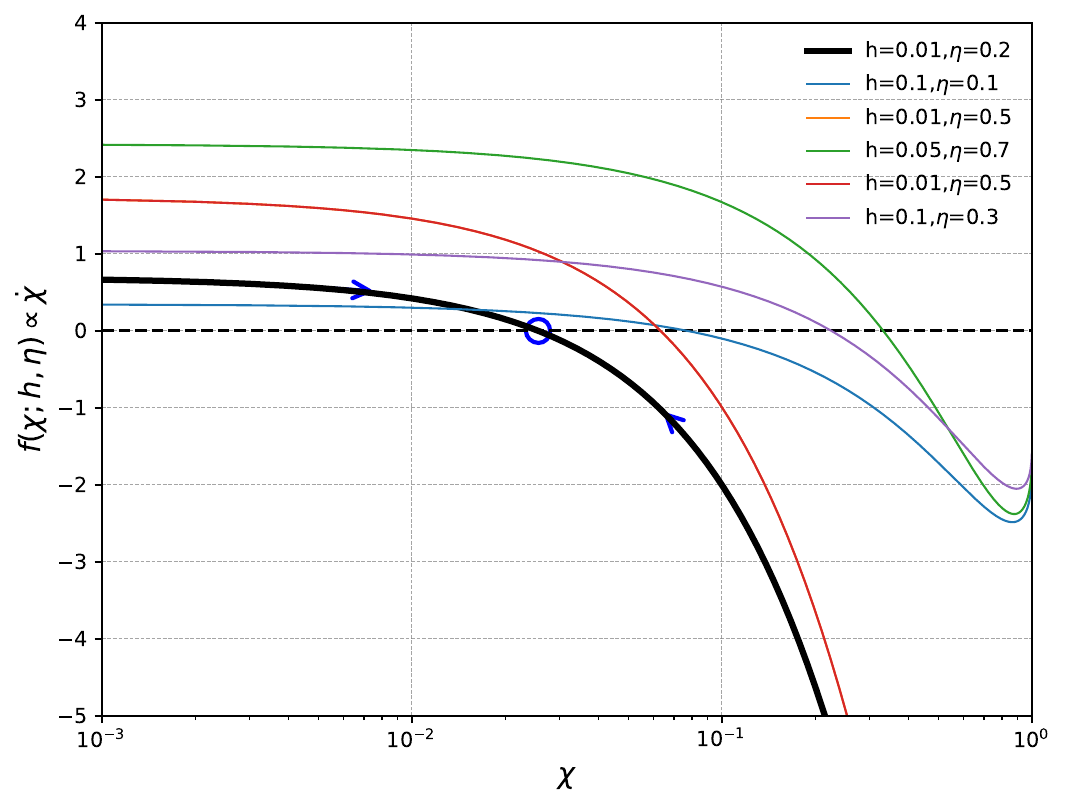}
    \caption{The function $f(\chi; h, \eta)$ for different values of $h$ and $\eta$. The dashed line indicates the zero line, where the equilibrium spin $\chi_{\rm eq}$ is located. The intersection points between the curves and the zero line indicate the equilibrium spin for different values of $h$ and $\eta$.}
    \label{fig:fx}
\end{figure}

For a BH gaining angular momentum via accretion and losing angular momentum through the BZ process, the evolution of the $\jbh$ is given by \cite{1972ApJ...178..347B,1998MNRAS.294..667W} (adopting the natural units where $c=G=1$):
\begin{equation}
\frac{d\jbh}{dt} =\Xi\dot{M}_{\rm rest}-\frac{\Pbz}{\Omega_{\rm F}},
\label{eq:2}
\end{equation}
where $\Xi$ is the specific angular momentum of the accreted matter, which depends on $\mbh$ (the BH mass) and $\chi$ (dimensionless spin of BH) (see {\bf Methods: The derivation of the universal equilibrium spin}), $\dot{M}_{\rm rest}$ is the accretion rate of the rest mass from the accretion flow, $\Omega_{\rm F}$ is the angular velocity of the magnetic field lines penetrating the BH event horizon, and $\Pbz$ is the power extracted via the BZ process. Conversely, the energy of the BH also evolves as accreted matter carries energy and the BZ process extracts energy from the BH. The evolution of the BH mass is given by \cite{1998MNRAS.294..667W},
\begin{equation}
    \dot{\mbh}=\epsilon\dot{M}_{\rm rest}-\Pbz,
    \label{eq:3}
\end{equation}
where $\epsilon$ is the specific energy of the accreted matter, which is independent of the BH mass, but only on the $\chi$ parameter (see {\bf Methods: The derivation of the universal equilibrium spin}). 
The power of BZ process is given by \cite{1999ApJ...523L...7A}:
\begin{equation}
    \Pbz = \frac{1}{32} B^2 \Rh^2 \chi^2 \alpha (1 - \alpha),
    \label{eq:6}
\end{equation}
where $B$ is the magnetic field strength at the event horizon; $\Rh$ is the radius of the BH; and $\alpha$ is the ratio between the angular velocity of the magnetic field $\Omega_F$ and that of the BH event horizon $\Omega_{\bullet}$. For MAD accretion, where the ram pressure of the accretion flow is balanced by the magnetic pressure at the event horizon, we have:
\begin{equation}
\frac{B^2}{8\pi}=\frac{\dot{M}_{\rm rest}}{2\pi h\Rh^2},    
\label{eq:7}
\end{equation}
where $h$ is the height of the accretion flow at the event horizon. Combining equations (\ref{eq:2},\ref{eq:3},\ref{eq:6},\ref{eq:7}), we can derive the evolution of the spin parameter $\chi$ of the BH (see detailed derivation in the {\bf Methods}):
\begin{equation}
    \dot{\chi} = \frac{\dot{M}_{\rm rest}}{4 h \mbh} \left[-(1 - \alpha) \alpha \chi^3 + \chi \left((\alpha-1) \left(1 + \sqrt{1 - \chi^2}\right) - 8 h \epsilon(\chi)\right) + 4 h\xi(\chi, \eta)\right].
    \label{eq:9}
\end{equation}
In the above equation, $\xi(\chi, \eta)\equiv\Xi/\mbh$ demontes the specific angular momentum of the accreted matter in units of the BH mass, and $\eta$ is the degree of circularization of the accreted matter at the innermost stable circular orbit (ISCO) (see {\bf Methods}). From the above equation, we see that the time derivative of the spin parameter, $\dot{\chi}=0$, when the accretion rate $\dot{M}_{\rm rest}$ is zero, or in the non-trivial case, when the part in the square bracket, denoted as $f(\chi; \alpha, h, \eta)$, is zero. This means that for MAD accretion, the spin parameter $\chi$ attains an equilibrium value when the BZ process extracts energy from the BH at a rate balancing the angular momentum gain from the accretion flow. Note that $f(\chi; \alpha, h, \eta)$ is independent of the mass of the BH, or the initial spin parameter, depending soley on $\chi$, $\alpha$ and $h$. The equilibrium spin $\chi_{\rm eq}$ is the root of the equation:
\begin{equation}
    f(\chi_{\rm eq}; \alpha, h, \eta) =0.
\end{equation}
If we assume that interactions between the specific accretion flow and the BH always yield universal values of $\alpha$, $h$ and $\eta$, the the equilibrium spin will be a universal value, which is independent of the mass of the BH, the initial spin and the history of accretion. As established in the literature, $\alpha$ always self-adjusts to 0.5, the value at which $P_{\rm BZ}$ reaches its maximum \cite{1982MNRAS.198..345M,1986bhmp.book.....T}. We therefore fix $\alpha=0.5$ in the subsequent analyses. We plot the function $f(\chi; h, \eta)$ in Figure \ref{fig:fx} for different values of $h$ and $\eta$. Since $f(\chi; h, \eta)$ is proportional to $\dot{\chi}$, therefore, when $\chi$ is smaller than $\chi_{\rm eq}$, $\dot{\chi}$ is positive, which means that $\chi$ will increase (the blue arrow on the curve); when $\chi$ is larger than $\chi_{\rm eq}$, $\dot{\chi}$ is negative, which means that $\chi$ will decrease. Therefore, the equilibrium spin $\chi_{\rm eq}$ (indicated with the blue circle in figure \ref{fig:fx}) is a stable point of the evolution of the spin parameter $\chi$. 

In {\bf Methods} section, we further demonstrate that $\chi_{\rm eq}$ is independent of the accretion rate history and the initial spin and mass of the BH. Therefore, $\chi_{\rm eq}$ is a universal value for all remnant BHs from natal MAD accretion. A direct prediction from this scenario, we expect to find a population of stellar-mass BHs with this universal spin value. In the past 50 years, the spins of stellar-mass BH have been mainly measured via their X-ray emission from accretion disks (e.g., fitting of the continuum spectral energy density \cite{1997ApJ...482L.155Z,2014SSRv..183..295M}; Fe K$\alpha$ line profile \cite{1989MNRAS.238..729F,1995Natur.375..659T,2014SSRv..183..277R,2017MNRAS.467..145F} and quasi-periodic oscillations \cite{2017MNRAS.467..145F}). However, the spins of these BHs are likely altered by the long-term non-MAD state accretion process in the X-ray binary phase. Therefore, we do not expect their spin to be at the above mentioned universal value.

The merger of a pair of BHs emits GW, which are the main targets of the working terrestrial GW detectors LIGO-Virgo-KAGRA network \cite{2009RPPh...72g6901A,2015CQGra..32b4001A,2019NatAs...3...35K}. The spin of both BHs are encoded in the GW waveform, and thus can be inferred using the Bayesian approach. The latest GW transient catalogue (GWTC-4.0) contains 154 BH binary merger events (with the false-alarm rate (FAR)$<1$\,yr$^{-1}$)  \cite{2025arXiv250818082T}. For each event, the posterior samples of all the relevant parameters are released in the catalogue. We further employ the hirarchical Bayesian inference method to infer the population properties of the BBHs. Especially, we are interested in the distribution of the spins of the 2nd-born BHs, as we expect that the 2nd-born BH is more likely to be the remnant of a type II GRB, since the progenitor of the 2nd-born BH may have been tidally spun up by the 1st-born BH so that it has sufficient angular momentum to support an accretion disk in its natal collapse. Moreover, there is no further mass/angular momentum transfer after the formation of the 2nd-born BH to alter its natal spin (see discussion in details in the {\bf Method: Progenitor evolution channels}). We assume the secondary BH is the 2nd-born BH, although there could be pollution in the sample due to mass ratio reversal (see discussion in the {\bf Method: mass ratio reversal}).

During the hierarchical Bayesian inference, we in general follow the method and population models as described in \cite{2023PhRvX..13a1048A,2025arXiv250818083T}. We make some modifications to the spin distribution model of the 2nd-born BH: we assume that the spin distribution of the 2nd-born BH is a composed by a narrow Gaussian and a wide Gaussian (see {\bf Method: The evidence of equilibrium spin in a population of BH found in GWTC-4.0} for details). The inferred posterior distribution of the relevant hyper-parameters are shown in figure \ref{fig:corner}. We find that there is a dominant population of the 2nd-born BHs with a narrow spin distribution centered at $\sim0.05$ with a very narrow width (consistent with zero). The Bayesian factor for the existence of the above-mentioned narrow component is $453.47$, indicating strong evidence. The inferred underlying population model for the $\chi_2$ (hyper-parameter averaged) is plotted in figure \ref{fig:chi2distribution}.  This is consistent with the prediction of the universal equilibrium spin scenario for a certain combination of $h$ and $\eta$ (see figure \ref{fig:contour}), and the value is also very close to the value found in the recent numerical simulation \cite{2024ApJ...961..212J}. 
{\begin{figure}
    \centering
    \includegraphics[width=0.5\textwidth]{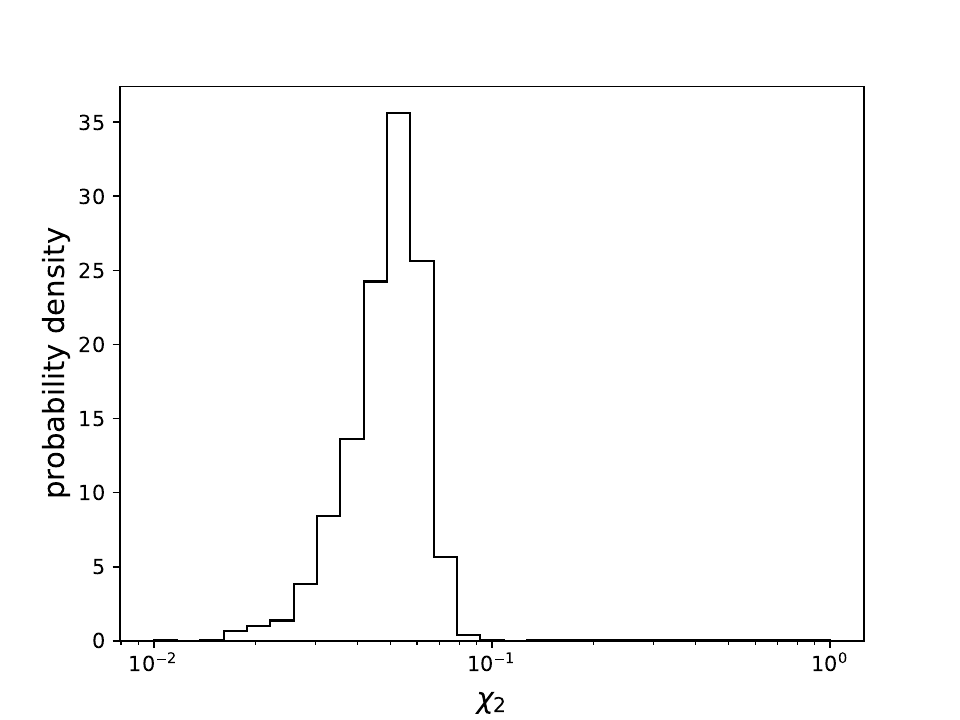}
    \caption{The inferred underlying population distribution of the spin parameter $\chi_2$ of the 2nd-born BHs in the BBH merger events in GWTC-4.0. The probability density function is averaged over the posterior distribution of the relevant hyper-parameters.}
    \label{fig:chi2distribution}
\end{figure}
We note that \cite{2025arXiv250818083T} also tried a truncated Gaussian spin distribution but the width is found to be wide at $\sim0.4$. We found that the reason that they did not reach a narrow spin distribution is due to the constraint on the numerical integration error during the treatment of the selection effect. We will discuss this in details in {\bf Methods: influence of the selection effect} section, and how we circumvent this problem in our treatment that allow us the inference of a narrow spin distribution. 

Since we demonstrate that an equilibrium spin is a natural outcome when there is BH spin energy extraction at its natal collapse, whereas such an equilibrium spin is not expected in an alternative scenario. Therefore, our findings provide direct evidence that the spin energy of BH can indeed be extracted by magnetic fields in a BZ-like process. Furthermore, since the equilibrium spin is function of the ratio between the angular velocity of the magnetic field and that of the BH event horizon, and the scale height of the accretion flow at the event horizon, the inferred $\mu_\chi$ can be used to constrain the accretion physics near the event horizon. Another implication from this work is that a dominating fraction of 2nd-born BHs in the binary BH (BBHs) merger sample have likely undergone the MAD accretion process at birth. This suggests that these BHs could serve as progenitors of gamma-ray bursts (GRBs). Consequently, at least a sub-population of type-II GRBs may originate from close binary systems, where the progenitor star of the 2nd-born BH is tidally spun up by its companion BH, facilitating the formation of an accretion disk during core collapse. 

\section{Methods}
\subsection{The derivation of the universal equilibrium spin}\label{sec:a}
Here we give details of the derivation of the universal equilibrium spin of a BH accreting in the MAD state while launching a BZ jet. In the case of a BH gaining angular momentum from accretion, and lossing angular momentum through the BZ process, the evolution of the $\jbh$ is given by equations (\ref{eq:2},\ref{eq:3}). Now note that the angular momentum of a Kerr BH is given by:
\begin{equation}
\jbh = \chi\mbh^2,
\label{eq:1}
\end{equation}
where $\chi$ is the dimensionless spin parameter in the Kerr metric. 
From equation \ref{eq:1}, we can see that,
\begin{equation}
    \dot{\jbh}=\dot{\chi}M^2+2\chi M\dot{M}_\bullet.
\end{equation}

Taking equations \ref{eq:2} and \ref{eq:3} into the above equation, we have:
\begin{equation}
    \dot{\chi} = \frac{-\Pbz + \mbh \dot{M}_{\rm rest} \Omega_{\rm F}\xi(\chi, \eta) - 2 \chi \mbh \Omega_{\rm F} \dot{\mbh}}{\mbh^2 \Omega_{\rm F}},
    \label{eq:4}
\end{equation}
where $\xi(\chi, \eta)\equiv\Xi/\mbh$ is the specific angular momentum of the accreted matter in units of the BH mass, which does not depend on the mass of the BH. The above equation can be further written as:
\begin{equation}
    \dot{\chi} = \frac{-2 \chi \mbh \Omega_{\rm F} \left( \epsilon(\chi) \dot{M}_{\rm rest} - \Pbz \right) - \Pbz + \mbh \dot{M}_{\rm rest} \Omega_{\rm F}\xi(\chi, \eta)}{\mbh^2 \Omega_{\rm F}}.
    \label{eq:5}
\end{equation}
In the above equations, The specific energy of the accreted matter is \cite{1972ApJ...178..347B}:
\begin{equation}
    \epsilon(\chi) = \frac{1 - 2 \kappa^{-2} + \chi \kappa^{-3}}{\sqrt{1 - 3 \kappa^{-2} + 2 \chi \kappa^{-3}}},
    \label{eq:10}
\end{equation}
and the specific angular momentum of the accreted matter divided by $\mbh$ is given by:
\begin{equation}
    \xi(\chi,\eta) = \eta\kappa \frac{1 - 2 \chi \kappa^{-3} + \chi^2 \kappa^{-4}}{\sqrt{1 - 3 \kappa^{-2} + 2 \chi \kappa^{-3}}}.
    \label{eq:11}
\end{equation}
The parameter $\eta$ is the fraction of the circularization of in-falling matter at the ISCO. When $\eta=1$, the matter is fully circularized at ISCO before being accreted by the BH. When $\eta=0$, the matter is radially in-falling without any angular momentum.  
In the above two equations, $\kappa\equiv\sqrt{R_{\rm ISCO}/\mbh}$, where $R_{\rm ISCO}$ is the radius of ISCO of a Kerr BH, which is given by:
\begin{equation}
    R_{\rm ISCO} = \mbh \left(3 + Z_2 - \sqrt{(3 - Z_1)(3 + Z_1 + 2 Z_2)}\right),
\end{equation}
where
\begin{equation}
    Z_1 = 1 + (1 - \chi^2)^{1/3} \left[ (1 + \chi)^{1/3} + (1 - \chi)^{1/3} \right]
\end{equation}
and
\begin{equation}
    Z_2 = \sqrt{3\chi^2 + Z_1^2}.
\end{equation}
As we can see, $\kappa$ is a function of the spin parameter $\chi$ only, independent of the mass of the BH. Therefore, the specific energy ($\epsilon$) and specific angular momentum divided by $\mbh$ ($\xi$) are also independent of the mass of the BH, but only depend on the spin parameter $\chi$ (and the constant $\eta$ for $\xi$).

Considering the expression of the radius of BH event horizon:
\begin{equation}
    \Rh=\left(1+\sqrt{1-\chi^2}\right)\mbh,
\end{equation}
and the angular velocity of the BH event horizon:
\begin{equation}
    \Omega_{\bullet}=\frac{\chi}{2 \mbh \left(1 + \sqrt{1 - \chi^2}\right)},
\end{equation}
we can rewrite equation \ref{eq:6} as:
\begin{equation}
    \Pbz = \frac{1}{32} (1 - \alpha) \alpha B^2 \chi^2 \left(1 + \sqrt{1 - \chi^2}\right)^2 \mbh^2.
    \label{eq:7}
\end{equation}
Now, the $B$ and the accretion rate $\dot{M}_{\rm rest}$ are related through the MAD condition in equation \ref{eq:8}. Taking the expression of $\Rh$ into equation \ref{eq:8}, we have:
\begin{equation}
    B^2 = \frac{4 \dot{M}_{\rm rest}}{h \left(1 + \sqrt{1 - \chi^2}\right)^2 \mbh^2}.
    \label{eq:8}
\end{equation}
Now, we can substitute equation \ref{eq:8} into equation \ref{eq:7}, and equation \ref{eq:7} further into equation \ref{eq:5}, we can obtain the evolution of the spin parameter $\chi$ as shown in equation \ref{eq:9}. 

As shown in the main text, the equilibrium spin $\chi_{\rm eq}$ is the root of the function $f(\chi; \alpha, h, \eta)$, and therefore it is a function of $h$ and $\eta$ (since we have fixed $\alpha=0.5$). We plot the contour of $\chi_{\rm eq}$ as a function of $h$ and $\eta$ in figure \ref{fig:contour}.
\begin{figure}
    \centering
    \includegraphics[width=\textwidth]{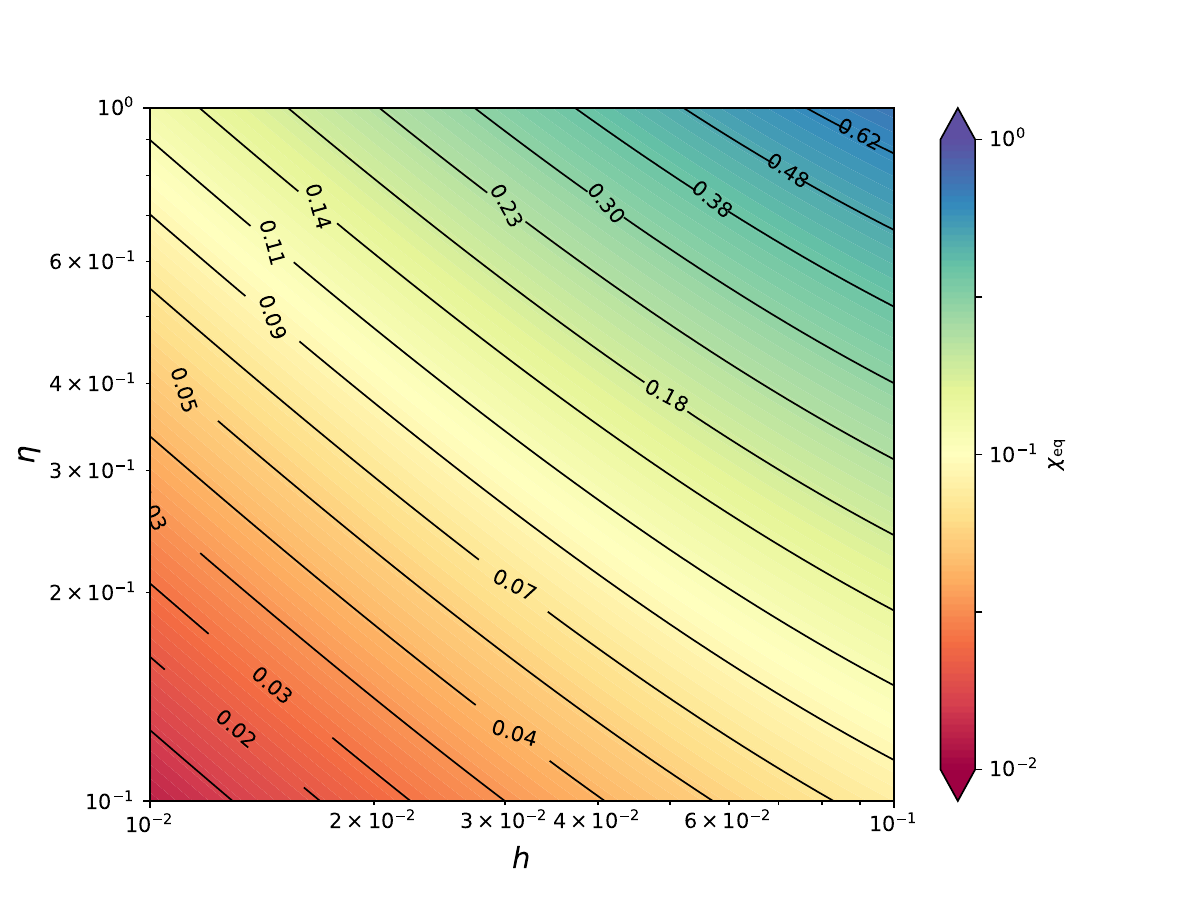}
    \caption{The contour plot of the equilibrium spin $\chi_{\rm eq}$ as a function of $h$ and $\eta$.}
    \label{fig:contour}
\end{figure}
\subsection{The evolution of the mass and spin of the BH under MAD accretion and the BZ process}
The time derivative of the mass of the BH is given by equation \ref{eq:3}. Substituting equation \ref{eq:7} into equation \ref{eq:3}, we have:
\begin{equation}
    \dot{\mbh} = \left(-\frac{(1 - \alpha) \alpha \chi^2}{8 h} + \epsilon(\chi)\right) \dot{M}_{\rm rest}.
    \label{eq:12}
\end{equation}

Combining equations \ref{eq:9} and \ref{eq:12}, and assuming a rest-mass accretion rate $\dot{M}_{\rm rest}(t)$ as a function of time, we can solve the evolution of the mass and spin of the BH simultaneously. 

As a fiducial scenario, we assume a power law accretion rate as: 
\begin{equation}
   \dot{M}_{\rm rest}=\dot{M}_0\left(\frac{t+\tau}{\tau}\right)^{-\beta},
\end{equation} 
when $\beta=5/3$, corresponding to the well-known value given by \cite{1988Natur.333..523R} in tidal disruption event. Other values like $\beta=19/16$ and $\beta=1.25$ are also employed in different physical scenarios \cite{2017ApJ...844..114Y}. The normalization factor $\dot{M}_0$ can be related to the total accreted matter as: 
$\dot{M}_{\rm rest}=\frac{(\beta-1)M_{\rm tot}}{\tau\beta}.$
We randomly choose the values of $M_0$ from a uniform distribution between 3 and 5 $M_\odot$, $\chi_0$ from a uniform distribution between 0.0 and 0.997, $M_{\rm tot}$ from a uniform distribution between 5.0 and 15.0 $M_{\rm tot}$, $\beta$ from a uniform distribution between 1.0 and 2.0, and $\tau$ from a log-uniform distribution between 100 and 1000 s. The resulted evolution of $\chi(t)$ and $\mbh(t)$ are shown in figure \ref{fig:chi_mbh}. 

\begin{figure}[h]
    \centering
    \includegraphics[width=0.95\textwidth]{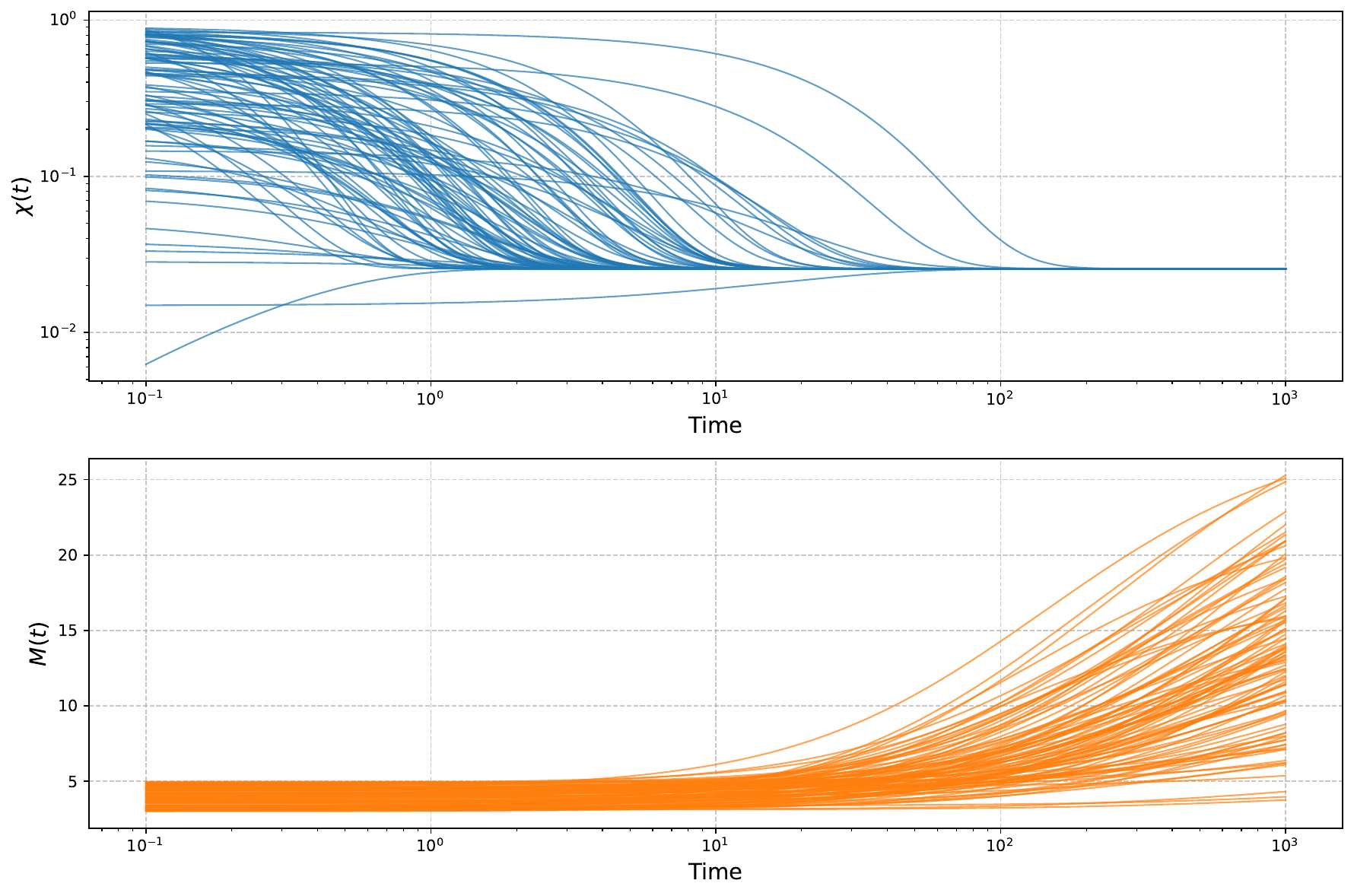}
    \caption{The evolution of the spin parameter $\chi$ (upper panel) and the mass of the BH $\mbh$ (lower panel) for a population of BHs with different initial conditions. }
    \label{fig:chi_mbh}
\end{figure}

As we can see, within reasonable initial $M_0$ and mass accretion scenario, the equilibrium spin $\chi_{\rm eq}$ is reached within tens of seconds to hundreds of seconds. It means a population of BHs which underwent the MAD accretion phase at their formation will have a natal spin at the universal value of $\chi_{\rm eq}$, which is independent of the mass of the BH, the initial spin of the seed BH (from the prompt-collapsed stellar core) and the history of the accretion. 
\subsection{Progenitor evolution channels}
It is believed that dominating population of the BBH merger events detected by LVK are form through the isolated binary massive star evolution channel: the progenitors of the BBH are massive star, which are initially separated in distance of $>R_\odot$. The primary star evolves faster and first collapses into a BH. The later evolution of the secondary star leads either to a common envelope phase \cite{2012ApJ...759...52D,2014A&A...564A.134M,2016Natur.534..512B,2016MNRAS.462.3302E,2018MNRAS.479.4391M,2018A&A...619A..77K,2018MNRAS.481.1908K,2020ApJ...898...71B,2021A&A...645A..54K}, or a stable mass transfer phase \cite{2017MNRAS.471.4256V,2021A&A...650A.107M}, which will cause the orbit to decay significantly. In both cases, the natal spin of the 1st born BH will be altered. Therefore, we do not expect the spin of the 1st born BH to be at the equilibrium spin value we found above. 

On the other hand, the progenitor of the 2nd born BH (will remain a stripped star after then mass transfer phase) may have been tidally spun up by the 1st born BH, when the separate is close enough \cite{2016MNRAS.462..844K,2017MNRAS.467.2146K,2017arXiv170708978H,2018MNRAS.473.4174Z,2018A&A...616A..28Q}. This process can effectively increase the angular momentum of the progenitor star, so that it has sufficient angular momentum to support an accretion disk in its natal collapse. Moreover, there is no further mass/angular momentum transfer after the formation of the 2nd born BH to alter its natal spin. Therefore, we expect that the spin of the 2nd born BH is more likely to be at the equilibrium spin value we found above.

Except the above mentioned channels, there are other proposed channels to form BBHs. For example, in the chemically-homogeneous evolution channel \cite{2016MNRAS.458.2634M,2016MNRAS.460.3545D,2016A&A...588A..50M}, the two low-metallicity massive stars are initially in an already close orbit, and they are tidally locked during their whole evolution. In this case, both stars are expected to be spun up effectively by the companion star, and they evolves chemically homogeneously avoiding a Roche-lobe overflow. As a result, there is no mass transfer expected between the two stars, and thus both BHs could be expected to be born with the equilibrium spin. 

In the dynamical formation channel \cite{2016ApJ...824L...8R,2017MNRAS.464L..36A,2017ApJ...840L..14S}, BBHs are formed through dynamical interactions in dense stellar environments like globular clusters or nuclear star clusters. In this cases, both BHs might have undergone hierarchical mergers \cite{2016ApJ...816...65A,2017ApJ...841...77A,2019MNRAS.486.4781F}, or the progenitors underwent direct collapse without enough angular momentum to support an MAD accretion disk during their formation. As a result, their spins are not expected to be at the equilibrium spin value.

There is also a proposed channel where BBHs are formed and merge in active galactic nuclei (AGN) disks \cite{1999ApJ...521..502C,2016ApJ...831..187A,2017MNRAS.464..946S,2018ApJ...859L..25Y,2018ApJ...866...66M,2021ApJ...916L..17W,2023PhRvD.107f3007L,2024ApJ...969...37L}. In this case, both BHs may undergo significant gas accretion from the AGN disk before their merger, and thus their spins are expected to be aligned and significantly increased from their natal values. 

\begin{figure}
    \centering
    \includegraphics[width=0.5\textwidth]{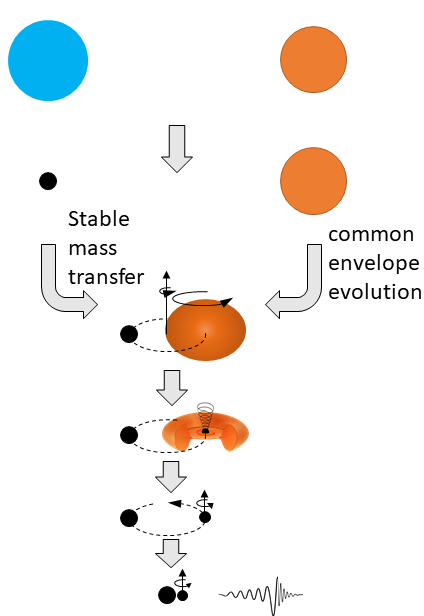}
    \caption{A schematic diagram of the evolution channel of the BBH system leading to the formation of a 2nd born BH with equilibrium spin.}
    \label{fig:schematic}  
\end{figure}

\subsection{The evidence of equilibrium spin in a population of BH found in GWTC-4.0}
In the event of a merger of a pair of BHs, the parameters of the binary system are encoded in the GW waveform emitted from it. Those parameters include those internal of the binary, including the masses and spins of both BHs, and the external parameters, including the sky location, distance, inclination angle, polarization angle, and the merger time. The probability distribution of those parameters can be inferred using the Bayesian approach:
\begin{equation}
    p(\Vec{\Theta}|h_{\rm data})\propto p(\Vec{\Theta})\mathcal{L}\left(h_{\rm data}|h_{\rm theory}(\Vec{\Theta})\right), 
\end{equation}
where $\Vec{\Theta}$ is the set of parameters of the event, $h_{\rm data}$ is the observed GW data, $h_{\rm theory}(\Vec{\Theta})$ is the theoretical GW waveform given the parameters $\Vec{\Theta}$, and $p(\Vec{\Theta})$ is the prior distribution of the parameters. As mentioned above, the posterior distribution $p(\Vec{\Theta}_i|h_{\rm data})$ of each GW event has been worked out and the MCMC samples are public available in \textit{Zenodo} \cite{ligo_virgo_kagra_collaboration_2025_10.5281zenodo17014085, ligo_virgo_kagra_collaboration_2022_10.5281zenodo5546663, ligo_virgo_collaboration_2022_10.5281zenodo6513631}. In total, we use $N_{\rm event}=154$ BBH events with a false-alarm rate below $1~{\rm yr}^{-1}$.

To investigate the population properties of BBHs, we employ a hierarchical Bayesian inference framework. In this approach, the event-level parameters $\Vec{\Theta}$ are assumed to be drawn from an underlying population distribution $p_{\rm pop}(\Vec{\Theta}|\Vec{\Lambda})$, which depends on a set of hyperparameters $\Vec{\Lambda}$. Given the observed data $\{h_{\rm data}\}$ for all events, the joint posterior distribution of $\Vec{\Lambda}$ and ${\Vec{\Theta}}$ is given by \cite{mandel2019extracting, vitale2022inferring, loredo2004accounting}:
\begin{equation}
\mathcal{L}(\{h_{\rm data}\}|\Vec{\Lambda},\{\Vec{\Theta}\})=p(\Vec{\Lambda},\{\Vec{\Theta}\}|\{h_{\rm data}\}) =
\frac{p(\Vec{\Lambda})}{\xi(\Vec{\Lambda})^{N_{\rm event}}}
\prod_{i=1}^{N_{\rm event}}
p(\Vec{\Theta}_i|h_{\rm data})
\frac{p_{\rm pop}(\Vec{\Theta}_i|\Vec{\Lambda})}{p(\Vec{\Theta}_i)}\,,
\label{eq:likelihoodfunction}
\end{equation}
where $p(\Vec{\Lambda})$ is the prior on the hyperparameters, and the term $\xi(\Vec{\Lambda})$, representing the fraction of detected events within the targeted population, corrects for selection effects by quantifying how the observed sample is biased by the parameter-dependent detection probabilities of sources. 
We assume the population model for $\chi_2$ is an overlap of two Gaussian distributions, corresponding to a narrow and a wide population. The means and standard deviations of the two populations are given by $\mu_{\chi,\rm narrow}$/$\mu_{\chi,\rm wide}$ and $\sigma_{\chi,\rm narrow} (<0.5)$/$\sigma_{\chi,\rm wide}(>0.5)$ respectively, and $\eta_{\chi_2, \rm narrow}$ is the ratio between these two populations. The formulation of the distribution of the rest of the parameters, i.e., the spin of the first - born BH, masses, and redshift distribution, is adopted as follows: the absolute value of the first - born BH $\chi_1$ follows a truncated Gaussian model, and the distribution of $\theta_1$, which is the angle between the first - born spin vector and the orbital angular momentum, is a combination of an isotropic and a truncated Gaussian model, the same as the default model in \cite{2025arXiv250818082T}. The distribution of $\theta_2$ is assumed to be identical as that of $\theta_1$. The redshift and masses distribution is adopted from the formulation of \cite{2023PhRvX..13a1048A}. The population model for different parameters, and the prior of the corresponding hyper-parameters are list in Table \ref{tab:parameters}:

\begin{table}
\centering
\renewcommand{\arraystretch}{1.5}
\begin{tabular}{c >{\centering\arraybackslash}p{6.5cm} c c c}
\hline\hline
\makecell{\textbf{physical}\\\textbf{parameter}} & \makecell{\textbf{population}\\\textbf{model}} & \makecell{\textbf{hyper-}\\\textbf{parameters}} & \textbf{prior} & \textbf{ref.} \\ 
\hline\hline
$z$ & $\frac{1}{(1+z)}\frac{\mathrm{d} V_c}{\mathrm{d} z}(1+z)^\kappa_z$ & $\kappa_z$ & Uniform(0,10) & \cite{2023PhRvX..13a1048A} \\
\hline

\multirow{7}{*}{$m_1$} 
& \multirow{7}{*}{\makecell{
$\big[(1-\lambda_p)\mathcal{P}(m_1|-\alpha_m, m_h)$\\
$+\lambda_p \mathcal{G}(m_1|\mu_m, \sigma_m)\big]\mathcal{S}(m_1|m_l, \Delta_m)$
}} 
& $\lambda_p$ & Uniform(0.001,0.99) & \multirow{7}{*}{\cite{2023PhRvX..13a1048A}} \\
 &  & $\alpha_m$ & Uniform(-4,12) &  \\
 &  & $\Delta_m$ & Uniform(30,110) &  \\
 &  & $m_l$ & Uniform(0.001,10) &  \\
 &  & $m_h$ & Uniform(0.05,10) &  \\
 &  & $\mu_m$ & Uniform(20,50) &  \\
 &  & $\sigma_m$ & Uniform(1,10) &  \\
 \hline

 \multirow{3}{*}{$m_2$} 
& \multirow{3}{*}{
$\mathcal{P}(m_2|\beta_m)\mathcal{S}(m_1|m_l, \Delta_m)$
} & $\beta_m$ & Uniform(-4,12) & \multirow{3}{*}{\cite{2023PhRvX..13a1048A}} \\
 &  & $\Delta_m$ & Uniform(30,110) &  \\
 &  & $m_l$ & Uniform(0.001,10) &  \\
 \hline

\multirow{2}{*}{$\chi_1$} 
& \multirow{2}{*}{\makecell{$\mathcal{T}(\chi_1|\mu_{\chi_1},\sigma_{\chi_1})$}}
& $\mu_{\chi_1}$ & Uniform(0,1) & \multirow{2}{*}{\cite{2025arXiv250818082T}} \\
 &  & $\sigma_{\chi_1}$ & Uniform(0,1) &  \\
 \hline

\multirow{2}{*}{$\cos{\theta_1}$} 
& \multirow{2}{*}{$\zeta_t\mathcal{T}(\cos{\theta_1}|1, \sigma_t)+(1-\zeta_t)$} & $\zeta_t$ & Uniform(0,1) & \multirow{2}{*}{\cite{2025arXiv250818082T}} \\
 &  & $\sigma_{\chi_1}$ & Uniform(0,1) &  \\
 \hline

\multirow{5}{*}{$\chi_2$} 
& \multirow{5}{*}{\makecell{
$\eta_{\chi_2, \rm narrow}\mathcal{T}(\chi_2|\mu_{\chi_2, \rm narrow},\sigma_{\chi_2, \rm narrow})$ \\
$+(1-\eta_{\chi_2, \rm narrow})\mathcal{T}(\chi_2|\mu_{\chi_2, \rm wide},\sigma_{\chi_2, \rm wide})$
}}
& $\mu_{\chi_2, \rm narrow}$ & Uniform(0,1) & \multirow{5}{*}{} \\
 &  & $\sigma_{\chi_2, \rm narrow}$ & Uniform(0,0.5) &  \\
 &  & $\mu_{\chi_2, \rm wide}$ & Uniform(0,1) &  \\
 &  & $\sigma_{\chi_2, \rm wide}$ & Uniform(0.5,1) &  \\
 &  & $\eta_{\chi_2, \rm narrow}$ & Uniform(0,1) &  \\
 \hline

$\cos{\theta_2}$
& Identical with $\cos{\theta_1}$ &  &  &  \\
\hline\hline
\end{tabular}

\caption{Summary of the population models, corresponding hyper-parameters, and adopted priors: In population model, $\frac{\mathrm{d} V_c}{\mathrm{d} z}$ is the differential comoving volume, $\mathcal{P}(m|-\alpha_m, m_h)$ is a normalized power-law distribution with spectral index $-\alpha$ and high-mass cutoff $m_h$, $\mathcal{G}(m|\mu_m, \sigma_m)$ is a normalized Gaussian distribution with mean $\mu_m$ and width $\sigma_m$, $\mathcal{S}(m|m_l, \Delta_m)$ is a smoothing function, which rises from 0 to 1 over the interval, $\mathcal{T}(\chi|\mu_{\chi}, \sigma_{\chi})$ is a normalized truncated Gaussian distribution with mean $\mu_{\chi}$, width $\sigma_{\chi}$ and cutoff at 0 and 1. The resulted posterior distribution of the hyper-parameters of the spin distribution of the 2nd-born BH is shown in figure \ref{fig:corner}.}
\label{tab:parameters}
\end{table}

\begin{figure}
    \centering
    \includegraphics[width=.8\textwidth]{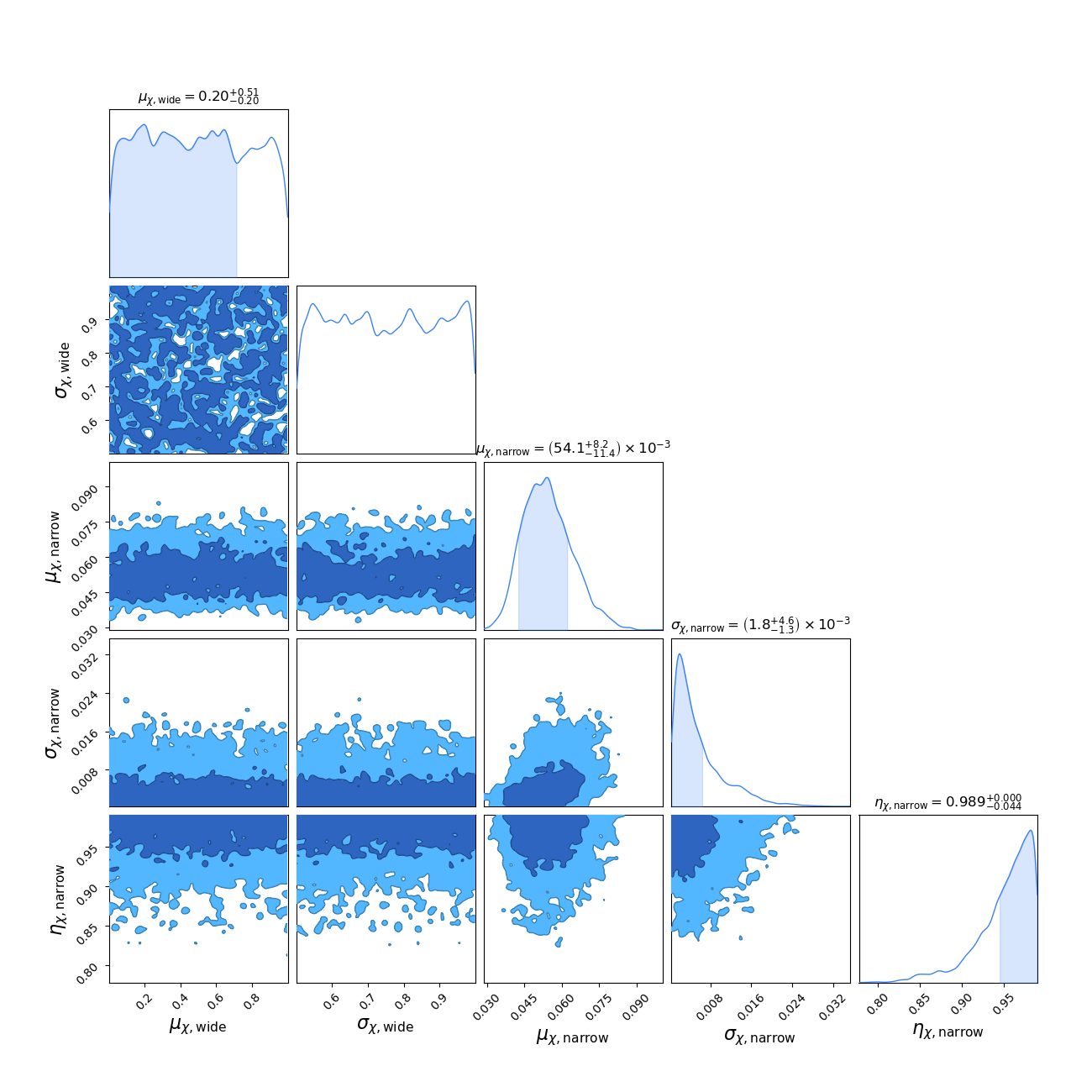}
    \caption{The corner plot of the posterior distribution of the relevant hyper-parameters of the spin distribution of the 2nd-born BH. The spin distribution is modeled as a composed of a narrow Gaussian and a wide Gaussian. The parameters $\mu_{\chi,{\rm narrow}}$ and $\sigma_{\chi,{\rm narrow}}$ are the mean and standard deviation of the narrow Gaussian component, respectively; $\eta_\chi$ is the fraction of the narrow Gaussian component in the total distribution.}
    \label{fig:corner}
\end{figure}


\subsection{Selection effect}
In the likelihood function in equation \ref{eq:likelihoodfunction}, the term $\xi(\Vec{\Lambda})$ is to correct for the selection effect bias, which represents the fraction of sources from the underlying population that are detectable by the observational pipeline. By definition, it is calculated by:
\begin{equation}
    \xi(\Vec{\Lambda})=\int\mathrm{d}\Vec{\Theta}p({\rm det}|\Vec{\Theta})p_{\rm pop}(\Vec{\Theta}|\Vec{\Lambda})\,.
\end{equation}
When a source can be detected, $p({\rm det}|\Vec{\Theta})=1$, otherwise $p({\rm det}|\Vec{\Theta})=0$. In practice, it can be estimated using a reweighted Monte Carlo integration approach. Specifically, it should first generate a set of simulated sources $\Vec{\Theta}_i$ drawn from a nominal (injection) distribution $p_{\rm inj}(\Vec{\Theta})$, with a total number of $N_{\rm inj}$ injections. For each injected source, we determine whether it would be detected by applying the same detection criterion used in the real search—namely, a false-alarm rate below $1~{\rm yr}^{-1}$. Suppose that $N_{\rm det}$ sources satisfy this detection threshold. Then, the Monte Carlo estimator of $\xi(\Vec{\Lambda})$ are given by
\begin{equation}
\xi(\Vec{\Lambda}) = \frac{1}{N_{\rm inj}} \sum_{i=1}^{N_{\rm det}}
\frac{p_{\rm pop}(\Vec{\Theta}_i|\Vec{\Lambda})}{p_{\rm inj}(\Vec{\Theta}_i)}\,,
\label{eq:numerical_xi}
\end{equation}
where $p_{\rm inj}(\Vec{\Theta}_i)$ is the injection distribution used to generate the simulated sources. The LVK collaboration provides such an injection set on \textit{Zenodo} \cite{ligo_virgo_kagra_collaboration_2025_10.5281zenodo16740128}, containing $N_{\rm inj}=1{,}499{,}244$ total injections and $N_{\rm det}=287{,}654$ detections.

Since the Monte Carlo integration approach is used instead of an exact integration, statistical errors are inevitably introduced. The resulting uncertainty in the joint posterior distribution can be expressed as \cite{2025arXiv250818082T}:
\begin{equation}
    \sigma^2_{\ln{\mathcal{L}}}(\Vec{\Lambda})=N_{\rm event}^2\frac{\sigma_{\xi}^2(\Vec{\Lambda})}{\xi(\Vec{\Lambda})^2}\,.
\end{equation}
The conventional threshold for ensuring the reliability of the inference is to require that $\sigma^2_{\ln{\mathcal{L}}}(\Vec{\Lambda})<1$ \cite{talbot2023growing}.

A potential issue arising from this criterion is that if the intrinsic population is largely different from the injection distribution, it can naturally lead to a large variance in the estimation of $\xi(\Vec{\Lambda})$ by equation (\ref{eq:numerical_xi}). In particular, the injected spin distribution is relatively wide. Therefore, imposing the condition $\sigma^2_{\ln{\mathcal{L}}}(\Vec{\Lambda})<1$ would therefore exclude scenarios where the spins are narrowly distributed. To examine this effect, we fix the hyper-parameters to the best-fit values obtained from the GWTC-4.0 sampling and compute $\sigma^2_{\ln{\mathcal{L}}}(\Vec{\Lambda})$ as a function of the variance $\sigma_{\chi}$ of an identical Gaussian spin distribution. The result is shown in Fig.~\ref{fig:sel_eff}. As seen in the figure, the spin variance is constrained to $\sigma_{\chi}>0.35$, consistent with the spin-distribution constraint reported in \cite{2025arXiv250818083T} (see Fig.~17 therein). Note that in \cite{2025arXiv250818083T}, the hard cut $\sigma^2_{\ln{\mathcal{L}}}(\Vec{\Lambda})<1$ is replaced with a smooth transition, therefore the inferred lower edge of the inferred $\sigma_{\chi}$ is smooth rather than a sharp truncation. If the threshold for the variance in the log-likelihood is modified, or if the GWTC-3 criterion—requiring the effective number of independent samples $N_{\rm eff}$ to be larger than $4N_{\rm events}$ \cite{farr2019accuracy}—is adopted instead, the inferred variance of the spin distribution will vary accordingly. In the appendix of M.~Mancarella \textit{et al.} \cite{2025PhRvD.111j3012M}, it is further demonstrated that adopting a more relaxed convergence criterion leads to smaller allowed spin variances.

\begin{figure}
\centering
\includegraphics[width=0.8\linewidth]{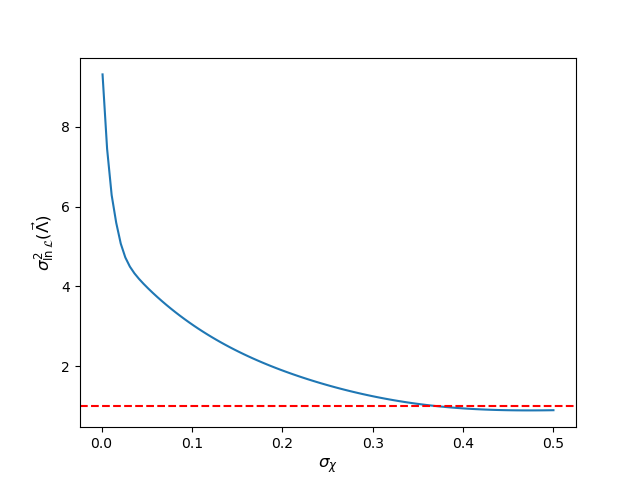}
\caption{Dependence of the variance of the log-likelihood on the variance $\sigma_{\chi}$ of an identical Gaussian spin distribution.}
\label{fig:sel_eff}
\end{figure}

One possible way to mitigate this bias is to increase the number of injected samples. However, since $\sigma^2_{\ln{\mathcal{L}}}\sim 1/N_{\rm inj}$, covering the full range of spin variances would require increasing $N_{\rm inj}$ by 4–5 orders of magnitude. From a computational and time-cost perspective, this approach is impractical. In addition, compared to the mass and distance parameters, the spin has a relatively minor impact on the SNR of GW detections and thus does not play a dominant role in the selection effects. We show in Fig.~\ref{fig:chi_snr} the relative change in the signal-to-noise ratio (SNR), defined as $\Delta {\rm SNR}/{\rm SNR} = \big({\rm SNR}(\chi_2) - {\rm SNR}(0)\big) / {\rm SNR}(0)$, as a function of the secondary spin. The different curves correspond to different BBH systems, we randomly sampled the chirp mass from 10-100 $M_{\odot}$ and mass ratio from 0.01-0.99. It shows that SNR variation is generally below $\sim10\%$ with the change of $\chi_2$ from 0 to 1, and is around a few percent for a smaller spin range. Therefore, we argue that when accounting for the selection effects, it is reasonable to neglect the $\chi_2$ dependence:
\begin{equation}
    \xi(\Vec{\Lambda}) = \frac{1}{N_{\rm inj}} \sum_{i=1}^{N_{\rm det}} \frac{p_{\rm pop}(\Vec{\Theta}_i|\Vec{\Lambda})/p_{\rm pop}(\Vec{\Theta}_{\chi_2, i}|\Vec{\Lambda}_{\chi_2})}{p_{\rm inj}(\Vec{\Theta}_i)/p_{\rm inj}(\Vec{\Theta}_{\chi_2, i})}\,,
\end{equation}
By doing this, the inferred distribution of $\chi_2$ will not influence the numerical inaccuracy in the estimation of $\xi(\Vec{\Lambda})$, and hence allow us to explore the possibility of a narrow spin distribution. 

\begin{figure}
\centering
\includegraphics[width=0.8\linewidth]{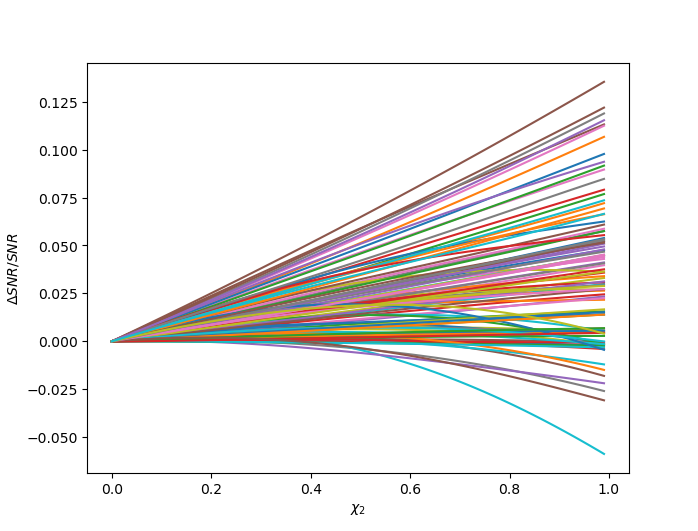}
\caption{Dependence of the relative change in the signal-to-noise ratio $\Delta {\rm SNR}/{\rm SNR}$ on the secondary spin $\chi_2$: Each curve corresponds to a different black-hole system with parameters randomly sampled from the underlying distribution, with chirp masses in the range 10-100 $M_{\odot}$ and mass ratios spanning 0.01-0.99. The SNR is computed using the IMRPhenomXPHM waveform template.}
\label{fig:chi_snr}
\end{figure}

\subsection{Mass Ratio Reversal}
There are some theoretical studies suggesting that in certain binary evolution scenarios, the secondary BH may end up being more massive than the primary BH, a phenomenon known as mass ratio reversal (MRR) \cite{2013PhRvD..87j4028G,2017MNRAS.471.2801S,2022ApJ...933...86Z,2022ApJ...938...45B}. In these cases, our target 2-nd born BH would correspond to the primary BH with $\chi_1$. In figure \ref{fig:mrr}, we show the posterior distribution of $\chi_1$ (upper half of the violin diagram) and $\chi_2$ (lower half of the violin diagram) of individual events in the GWTC-4.0 catalogue, together with a dashed line indicating the equilibrium spin at $\sim0.05$; If we find that the $90\%$ credible lower limit of $\chi_2$ is higher than the equilibrium spin, we conclude that in this event, the secondary BH is unlikely to be consistent with the 2-nd born BH with equilibrium spin. For those events, we then consider the possibility of MRR, and check whether the primary BH's spin $\chi_1$ is consistent with the equilibrium spin. If their $90\%$ credible lower limit of $\chi_1$ is lower than the equilibrium spin (coloured in orange), we consider this event might be a candidate of the mass ratio reversal. For those whose $90\%$ credible lower limit of $\chi_1$ is also higher than the equilibrium spin (coloured in red), we consider that this event might correspond to other formation channels which results in both BHs having high spins (see {\bf Method: Progenitor evolution channels} for details).

\begin{figure}
    \centering
    \includegraphics[width=\textwidth]{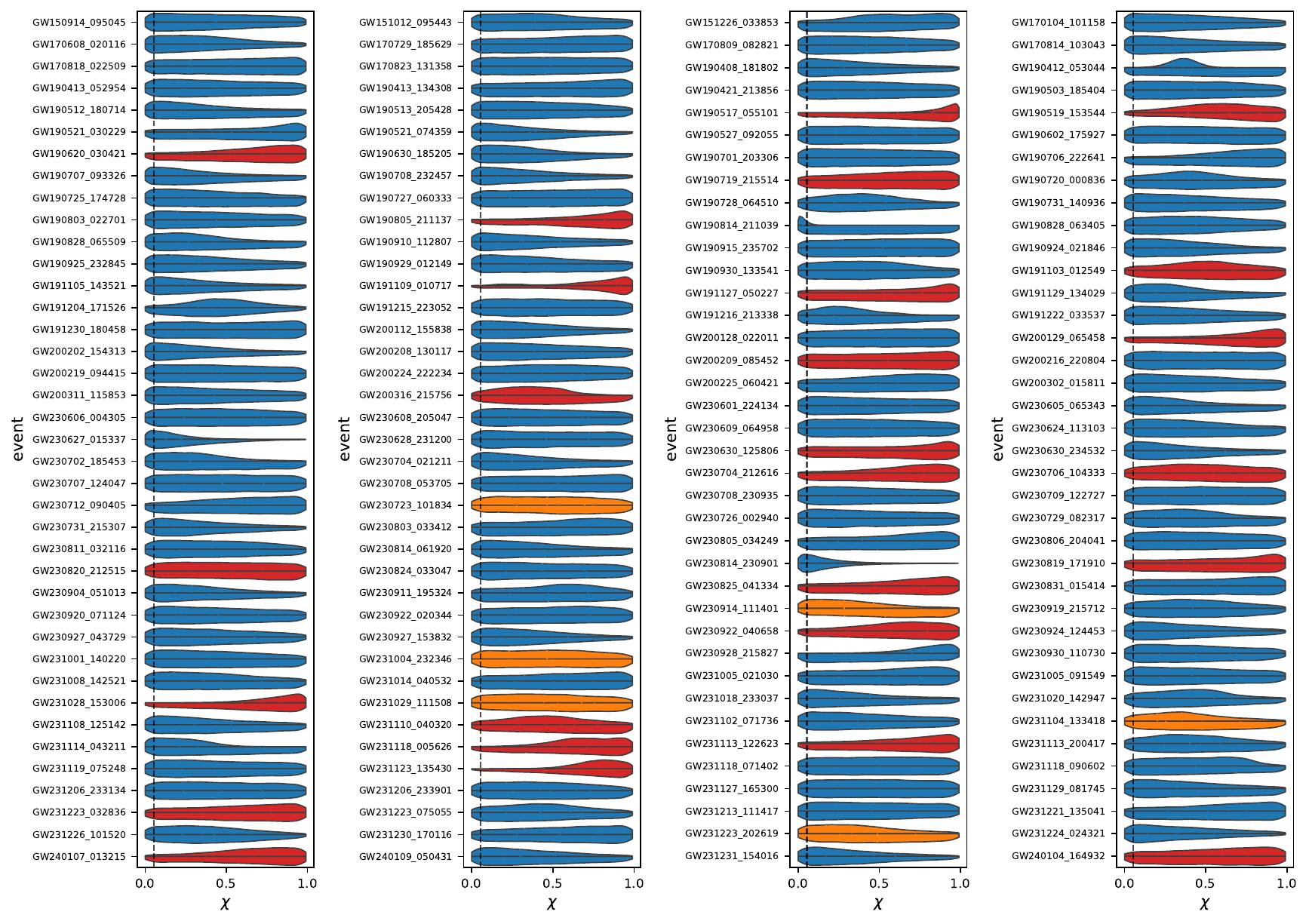}
    \caption{The posterior distribution of $\chi_1$ (upper half of the violin diagram) and $\chi_2$ (lower half of the violin diagram) of individual events in the GWTC-4.0 catalogue. The dashed line indicates the equilibrium spin at $\sim0.05$. Events coloured in orange are considered as candidates for mass ratio reversal, while those coloured in red are likely formed through other channels resulting in both BHs having high spins.}
    \label{fig:mrr}
\end{figure}

\section{Acknowledgements:} 
We benifit from the discussion with Profs Gijs Nelemans and Andrzej Zdziarski. This work is supported by the Chinese Academy of Sciences youth talent project under grant No. E32983U810 and China's Space Origins Exploration Program. 
\section{Reference list:}
\bibliography{main}{}
\bibliographystyle{naturemag}
\end{document}